# Observed Spacetime Dimensionality from Fundamental Principles

Rajat K. Pradhan*

Rajendra College, Balangir, Odisha, India-767002

(Date: 19.03.2013)

## Abstract

It is shown by very simple arguments that the observed 3+1 dimensionality of spacetime may be understood on the basis of four fundamental principles of physics namely, Causality, General Covariance, Gauge Invariance and Renormalizability. This approach not therefore only fixes the spacetime dimensionality at 3+1, but also imposes the compactification of extra dimensions in higher dimensional theories based on these principles such as string theory as an internal consistency requirement rather than an external imposition.

Key words: Causality, General Coordinate Invariance, Renormailizability, Gauge invariance

PACS: 11.10.Gh, 11.25Mj, 04.20Gz

---

*Email: rajat@iopb.res.in

## 1. Introduction

Ever since the advent of general relativity it has been increasingly realized that the structure of spacetime is inextricably linked with the 'objects' inhabiting it and their mutual interactions. Moreover as is well known, these 'objects' in it signify the structure i.e. the metric and topology of the base space, which is the observed large-scale spacetime, while their interactions can be understood as arising from extra small scale fibrations on it *a la* Kaluza-Klein. This unifying perspective has been the guiding force behind all higher dimensional theories when it comes to making contact with the observed world. What concerns us here is the determination of the base space dimensionality from basic principles which are at the heart of all current theories so that their very adoption would demand compactification as an internal consistency requirement rather than an adhoc imposition.

In this work, we seek to establish that the observed large-scale spacetime dimensionality can be determined by the dynamical theories based on just four fundamental principles viz.

(i) The Principle of Causality

(ii) The Principle of General Covariance

(iii) The Gauge Principle, and

(iv) The Principle of Renormalizability.

These four principles summarize our deepest understanding of nature so far in the sense that they are sufficient to lead us to a standard model-like framework for particles (fields) and their interactions (principles i, ii and iv) in a background spacetime determined by general relativity (i and ii). In this work we will take general relativity as a classical theory since we are interested in the large scale structure of spacetime. In what follows we start with an m+n spacetime and proceed to analyze the various restrictions on m and n following from these principles.

## 2.1 Restrictions from Causality

The requirement of causality imposes a severe restriction on spacetime dimensionality as it rules out multiple time dimensions. The primary reason for this is that multiple times inevitably entail the existence of tachyons alongside bradyons [1, 2], thereby leading to inconsistencies associated with non-causal events. We note that causality is the fundamental principle behind the requirement of sufficient predictability and stability as discussed by Tegmark [3] following the arguments of Dorling [4]. Further, the requirement of micro-causality is also fundamental to quantum field theories for the imposition of appropriate commutation or anti-commutation relations for Bosonic and Fermionic fields respectively. As is well known the existence of closed time-like curves (CTCs) does not allow causal quantum fields to be defined in spacetimes with multiple time dimensions. Causality therefore effectively rules out multiple time dimensions restricts m+n dimensions to only m+1 spacetimes.

## 2.2 Restrictions from General Coordinate Invariance (GCI)

In addition to leading us to general relativity, the principle of general Coordinate invariance also brings in the question of dimensions other than 3+1, since the range of the tensor indices is not fixed a priori by the theory and the tensor equations of physics could be regarded as valid in arbitrary spacetimes. However, for infra-space dimensions (m<3, n=1), it does have certain implications which can be used to rule out such spacetimes. As is well known, general relativity is trivial in 1+1 dimensions leading to the absence of gravity [5] while in 2+1 dimensions the curvature and Einstein tensors are related by

$$R^{\mu\nu}{}_{\alpha\beta} = \epsilon^{\mu\nu\sigma}{}_{\alpha,,\lambda} \, G^{\lambda}{}_{\sigma} \qquad \ldots \ldots \ldots \quad (1)$$

making source-free regions flat ($G^{\lambda}{}_{\sigma} = 0$) and allowing for nontrivial structure only in regions with source

$$G^{\lambda}{}_{\sigma} = 8\pi G T^{\lambda}{}_{\sigma} \neq 0 \qquad \ldots \ldots \ldots \quad (2)$$

with G being the gravitational constant with inverse mass dimensions.

As was shown by Deser *et al* [6], although a material particle acquires a gravitational self-interaction, it introduces a conical deficit directly proportional to its mass in the global topology of the 2+1 spacetime. This in turn means that in a QFT defined in such a spacetime the conserved charges arising out of gauge symmetries cannot be consistently defined unless they are also related to topological invariants like the quantum numbers following from spacetime symmetries (such as energy, which happens to be just the Euler characteristic of the spatial two-surface). This makes both gravity and any gauge theory global as well as topological in character thereby robbing them of their vital non-trivialities. Thus since general relativity allows for intrinsic spacetime curvature independently of sources only in m≥3 dimensions, we conclude that gravity becomes truly operative only in m≥3 spacetimes.

## 2.3 The Gauge Principle

For our purposes the principle can be taken as stating that all non-gravitational interactions be described by local gauge invariant quantum field theories. By itself it does not impose any restrictions on the spacetime dimensions but it does restrict the kind of terms that can appear in the action. We exclude Gravity from the gauge domain because even if a correct quantum theory of gravity is discovered or formulated it would not presumably have any bearing on the macroscopic spacetime dimensionality, since it would inevitably be a theory of small length and high energy scales. Therefore gravity can be treated as a classical theory of spacetime providing the background arena on which quantum field theories of the gauge interactions are to be built. Moreover in any spacetime, as long as time is restricted to one dimensions only, gauge interactions can , I general be either short-range or long-range type depending on whether the corresponding gauge fields are massive or massless, which fact can be used to fix m on the basis of Renormalizability.

## 2.4 The Principle of Renormalizability

Renormalizability [7, 8] as a criterion for determining acceptable quantum field theories is unavoidable for our purposes in the sense that no other principle can so effectively and unambiguously rule out supra-space dimensions (m > 3, n = 1). That Renormalizability could play a decisive role in the determination of spacetime dimensionality was first pointed out by Tangherlini [9]. Although there could be many gauge theories in any dimension, we require specifically a renormalizable QED whose gauge quanta will play the fundamental role of light waves in the formulation of special as well as general relativity at low energies appropriate for the observation of macroscopic spacetime dimensions.

In m+1 dimensions the mass dimension of the coupling constant appearing with a generic interaction $\mathcal{L}_i$ containing $N_{fi}$ number of fields of type $f$ with dimensionality

$$D_f = \sigma_f + (m-1)/2 \qquad \ldots \ldots \ldots (3)$$

and $d_i$ derivatives is given by :

$$\gamma_i = m+1 - d_i \Sigma_f N_{fi} D_f \qquad \ldots \ldots \ldots (4)$$

where $\sigma_f$ can be interpreted as the spin of the field quanta in 3+1 dimensions barring subtleties arising from gauge invariance e.g. $\sigma_f = 0$ for massless vector fields under consideration. The degree of divergence of a diagram with $E_f$ external field lines of type $f$ and $V_i$ vertices of the interaction type $\mathcal{L}_i$ is given by

$$D = m+1 - \Sigma_f E_f D_f - \Sigma_i V_i \gamma_i \qquad \ldots \ldots \ldots (5)$$

This has an upper bound

$$D \leq m+1 - \Sigma_f E_f D_f \qquad \ldots \ldots \ldots (6)$$

provided, the interactions satisfy the Renormalizability condition

$$\gamma_i \geq 0 \qquad \ldots \ldots \ldots (7)$$

which means

$$d_i + \Sigma_f N_{fi} D_f \leq m+1 \qquad \ldots \ldots \ldots (8)$$

For QED interaction this immediately gives m ≤ 3. We note that the equality in eq. (7) implies Renormalizability, while $\gamma_i > 0$ implies super-renormalizability. This is as far as the long-range type QED interaction is concerned.

Now, coming to the case of short range interactions, where massive gauge bosons are involved the principle of Renormalizability necessitates the introduction of Higgs fields which themselves must have renormalizable self-interaction as achieved in the Standard Model. Taking a general $\phi^s$ kind of interaction, with s an integral power, and using $\sigma_f = 0$ as required for scalar fields in the eq. (8), we readily get

$$m+1 \leq 2s/(s-2) \qquad \ldots \ldots \ldots (9)$$

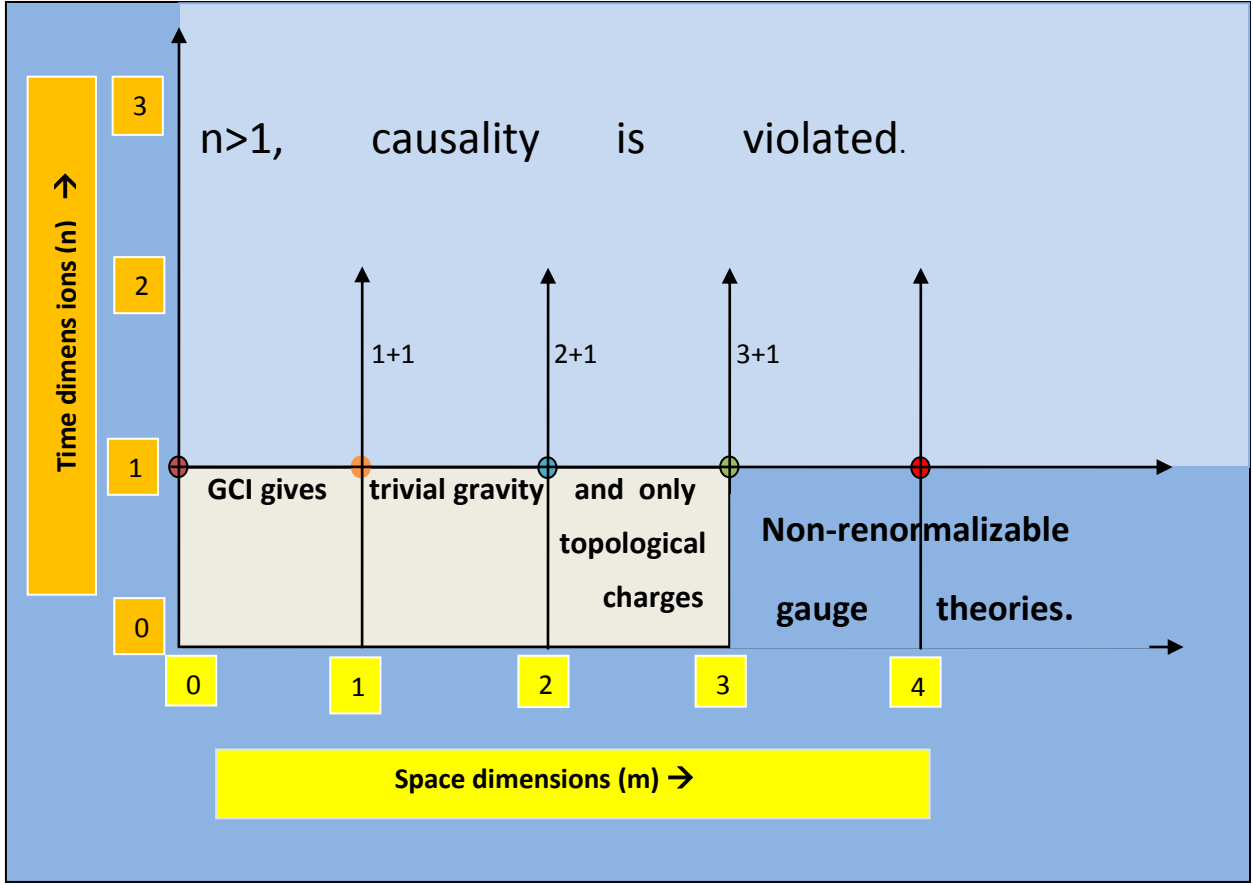

( **Fig. : spacetime dimensionality from fundamental principles**)

This tells us that the only renormalizable scalar field theories are those with $\phi^3$ and $\phi^4$ type interactions in m≤5 and m≤ 3 dimensions respectively. Of these, the $\phi^3$ theory is clearly unsuitable as a Higgs interaction and thus we are finally left with the only renormalizable choice:

$$s = 4, m \leq 3 \qquad \ldots \ldots \ldots (10)$$

Thus, the Gauge Principle along with the Principle of Renormalizability permits only m≤3 spacetimes, while general coordinate invariance demands that we can have nontrivial general relativity only for m≥3 spacetimes. Therefore we are unambiguously led to the observed 3+1 dimensionality of spacetime (Fig. 1 above), which we truly find ourselves in.

## 3. Conclusion

We have shown that the observed 3+1 macroscopic spacetime dimensionality can be understood on the basis of four fundamental principles. Further since for all higher dimensional theories like string theory and supergravity etc. the 3+1 dimensionality will be an internal consistency requirement rather than an external imposition since these four principles are the basic ones upon which these theories rest. It remains to be discussed in future work if we can reduce the number of fundamental principles required or whether we can get some of them from some more profound principle like the anthropic principle [10] or the holographic principle [11].

**Acknowledgements**

Most of the work was completed during my tenure as a doctoral scholar at the Institute of Physics, Bhubaneswar, where I benefited a lot from discussions with J. Maharana and J. Kamila. I express my gratitude to my teachers Mayadhar Sahu, Purandara Sahu, Chittaranjan Biswal and Suresh Nayak for many stimulating discussion sessions on various aspects.